\documentclass[manuscript,screen]{acmart}
\usepackage[colorinlistoftodos,prependcaption,textsize=small]{todonotes} 
\usepackage{tabulary} 
\usepackage{lscape}
\usepackage{tikz}
\usetikzlibrary {er, fit, bending}
\AtBeginDocument{%
  }

\setcopyright{acmlicensed}
\copyrightyear{2025}
\acmYear{2025}
\acmDOI{}

\acmConference[AIES 2025]{Eight AAAI/ACM Conference on AI, Ethics, and Society}
\acmISBN{}

\begin{document}

\title[A Taxonomy of Questions for Critical Reflection]{A Taxonomy of Questions for Critical Reflection in Machine-Assisted Decision-Making}

\author{Simon W.S. Fischer}
\email{simon.fischer@donders.ru.nl}
\orcid{0000-0003-2992-6563}
\affiliation{%
  \institution{Donders Institute for Brain, Cognition, and Behaviour}
  \city{Nijmegen}
  \country{Netherlands}
}

\author{Hanna Schraffenberger}
\affiliation{%
  \institution{iHub, Radboud University}
  \city{Nijmegen}
  \country{Netherlands}
}

\author{Serge Thill}
\affiliation{%
  \institution{Donders Institute for Brain, Cognition, and Behaviour}
  \city{Nijmegen}
  \country{Netherlands}
}

\author{Pim Haselager}
\affiliation{%
  \institution{Donders Institute for Brain, Cognition, and Behaviour}
  \city{Nijmegen}
  \country{Netherlands}
}

\renewcommand{\shortauthors}{Fischer et al.}

\begin{abstract}
    Decision-makers run the risk of relying too much on machine recommendations, which is associated with lower cognitive engagement. Reflection has been shown to increase cognitive engagement and improve critical thinking and therefore decision-making. Questions are a means to stimulate reflection, but there is a research gap regarding the systematic creation and use of relevant questions for machine-assisted decision-making. We therefore present a taxonomy of questions aimed at promoting reflection and cognitive engagement in order to stimulate a deliberate decision-making process. Our taxonomy builds on the Socratic questioning method and a question bank for explainable AI. As a starting point, we focus on clinical decision-making. Brief discussions with two medical and three educational researchers provide feedback on the relevance and expected benefits of our taxonomy.
    Our work contributes to research on mitigating overreliance in human-AI interactions and aims to support effective human oversight as required by the European AI Act.
\end{abstract}



\received{24 May 2025}
\received[accepted]{19 July 2025}
\received[revised]{11 August 2025}

\maketitle

\section{Introduction}

Reflection, also referred to as reflective practice or reflective thought, is an important skill in professional practices \cite{Schon1983}. It is beneficial in, for example, management \cite{Cunliffe2002}, law \cite{Casey2014}, and healthcare \cite{Mann2009}, and is part of many educational curricula \cite{Boud1998}.
As the American philosopher John \citet[p.9]{Dewey1933} states, reflection is an 
\begin{quote}
``Active, persistent, and careful consideration of any belief or supposed form of knowledge in the light of the grounds that support it and further conclusion to which it tends''.  
\end{quote}

As reflection helps to scrutinise the validity of assumptions and information \cite{Mezirow1990}, it has the potential to improve reasoning, judgement, and problem-solving \cite{Ghanizadeh2017, Khoshgoftar2023, Mamede2023}. Consequently, reflection has been shown to improve strategic decisions \cite{Walger2016} and diagnostic accuracy \cite{Mamede2008, Hess2015, Prakash2019}.
In addition, reflection enables decision-makers to recognise and articulate the reasons for their decisions, allowing them to direct their actions accordingly and take responsibility (forward-looking) and accountability (backward-looking). Responsible decision-making ideally leads to more desirable outcomes for all parties involved. 
In view of this, it seems sensible to encourage and support reflection during decision-making, especially as many high-impact decisions nowadays are informed by machine recommendations or predictions.

Decision-support systems (DSS) provide recommendations that assist decision-makers in domains like law, finance and healthcare. Physicians, for example, can use clinical DSSs to make a diagnosis or find a treatment option. Studies show, however, that operators tend to rely too much on these systems and accept incorrect recommendations \cite{Dratsch2023}, which is known as overreliance \cite{Passi2022}. 
In order to reduce harmful consequences of overreliance, the European AI Act \cite[\S14]{EuropeanParliamentandCounciloftheEuropeanUnion2024} requires:
\begin{quote}
    [...] that natural persons to whom human oversight is assigned are enabled, as appropriate and proportionate: [...]
    (b) to remain aware of the possible tendency of automatically relying or over-relying on the output produced by a high-risk AI system (automation bias), in particular for high-risk AI systems used to provide information or recommendations for decisions to be taken by natural persons;
\end{quote}

Various approaches therefore aim to mitigate overreliance by increasing the decision-maker's cognitive engagement through different interactions, such as presenting evidence for and against a decision \cite{Miller2023} or presenting explanations in the form of questions \cite{Danry2023}. Others argue that DSS should support deliberation and reflection rather than make recommendations \cite{Schmidt2023, Ma2024}. Similarly, in previous work we have introduced the idea of a \textit{reflection machine}, which asks questions to the decision-maker \cite{Cornelissen2022, Haselager2023, Fischer2025}. 
However, it remains unclear what kind of questions should be asked in order to effectively stimulate cognitive engagement and reflection.

To address this gap, we provide a taxonomy of questions that helps to systematically formulate relevant questions that stimulate critical reflection on the decision at hand, including the machine recommendation and the data that led to it. 
To identify relevant questions, the Socratic questioning method, named after the Greek philosopher Socrates, proves to be a useful starting point as it involves systematically asking questions to explore complex ideas, uncover assumptions, and gain a better understanding of the topic under investigation \cite{Paul2019}. As such, Socratic questioning has been shown to increase students' critical thinking \cite{Dalim2022, Ho2023}.
Although the Socratic method is slowly gaining traction in the field of human-machine interaction \cite{Lara2021,Ang2023,Liu2024b}, there is as yet no systematic approach for applying this questioning method to the context of machine-assisted decision-making. 

We therefore adapt a taxonomy of Socratic questions based on concepts of critical thinking \cite{Paul2019} to identify aspects to which questions should relate, and transfer it to the context of machine-assisted decision-making by synthesising it with a question bank for the design of human-centred explainable AI \cite{Liao2020, Liao2021, Liao2022}. We also draw inspiration, albeit only marginally, from a revised version of Bloom's taxonomy \cite{Krathwohl2002, Forehand2010} to identify cognitive processes to which questions can refer.

While we aim for a domain-independent taxonomy by building on general concepts of critical thinking, we take as a starting point clinical decision-making, where a physician uses a DSS to find a treatment option among several possibilities with uncertain outcomes for patients. To assess the transferability of our taxonomy, we apply it to the educational domain, where teachers use learning analytics dashboards to make decisions about students. We have discussed our taxonomy with medical and educational researchers, who have emphasised the relevance and expected benefits of our taxonomy. Nevertheless, an empirical evaluation of the effectiveness of certain questions in a specific decision-making context remains for future work. The contributions of this paper are as follows: 
\begin{itemize}
    \item With regard to mitigating overreliance (section \ref{dss}) and human-centred XAI (section \ref{human-xai}), we add to the growing literature on approaches to increase cognitive engagement, and, in particular, expand on existing proposals to promote reflection (section \ref{alternative-hci}).
    \item We discuss literature that shows that reflection increases cognitive engagement and reasoning, and thus decision-making (section \ref{reflection}), and that questions can stimulate reflection (section \ref{socratic-method}).
    \item We present a new taxonomy to help formulate relevant questions that stimulate reflection in machine-assisted decision-making (section \ref{taxonomy}).
    This taxonomy supports the design and development of interventions that go beyond machine recommendations and explanations. In addition, our taxonomy can benefit stakeholders such as developers, auditors, or consumers who need to remain critical in the development, evaluation, and use of DSS (section \ref{discussion}).
\end{itemize}

\section{Background}
\label{background}

In the following, we will discuss background work that led us to the creation of our question taxonomy. In order to sketch the problem of overreliance on machine recommendations and explanations (section \ref{dss}), we begin with a short review of the relevant literature on human-machine interaction, particularly in relation to the human factor in explainable AI (section \ref{human-xai}). From alternative interaction approaches, we derive that promoting \textit{cognitive engagement} is a promising method to reduce overreliance (section \ref{alternative-hci}). We identify the benefits of \textit{reflective thought} for cognitive engagement as it promotes critical thinking, reasoning, and thus responsible decision-making (section \ref{reflection}). Finally, we suggest that \textit{questions} can be useful to stimulate reflection and thus cognitive engagement, and introduce the Socratic questioning method, which is widely used in education (section \ref{socratic-method}).
In short, to arrive at our taxonomy, we combine relevant insights from empirical findings and literature on human-machine interaction, education, and philosophy.

\subsection{Overreliance on DSS and Explainable AI}
\label{dss}

Decision-support systems (DSS) provide recommendations based on information by using (simple) mathematical models, or more advanced machine learning models, including deep learning approaches. 
A meta-study of 106 experimental studies on machine-assisted decision-making found that decision-making accuracy generally decreases when decision-makers use DSS \cite{Vaccaro2024}. Decision-makers may tend to prefer machine recommendations over their own judgement \cite{Tschandl2020}, even when the recommendations are wrong \cite{Jacobs2021}. The acceptance of incorrect recommendations, referred to as overreliance, can be because of low vigilance or lack of cognitive engagement \cite{Passi2022}.

A common approach to mitigating overreliance is explainable AI (XAI) \cite{Vasconcelos2023, Longo2024}, which can, for example, provide information about which input was decisive for a particular output. It is assumed that these explanations will help the operator to assess the DSS recommendation and thus make an informed decision as to whether or not to follow the recommendation. 

The aforementioned meta-study, however, found that explanations do not lead to significant synergy effects in human-machine decision-making \cite{Vaccaro2024}. On the contrary, explanations may increase the likelihood of operators accepting a machine recommendation regardless of its correctness \cite{Bansal2021,Bussone2015,Ehsan2021}, thereby increasing overreliance \cite{Vered2023}. 
One reason for this is that the focus on numerical explanations can give a false sense of objectivity, leading operators, regardless of their expertise, to overestimate the capabilities of a DSS \cite{Ehsan2024a}. Consequently, providing additional information in the form of explanations does not automatically lead to more cognitive engagement of the decision-maker and to more accurate decisions \cite{Gajos2022}. 

\subsection{A Question Bank for XAI}
\label{human-xai}

To provide more relevant and actionable explanations, a question bank for explainable AI helps  designers and developers select the appropriate XAI technique that supports the operator's information needs \cite{Liao2020, Liao2021, Liao2022}. To this end, the question bank clusters and maps prototypical user questions to XAI techniques. Feature importance, for example, can provide insights into why-questions, such as ``Why was this prediction given?''.  

Answering the (technical) question of how a DSS arrived at a particular output is however not always desired or helpful. As one participant working on clinical decision-support systems mentioned in the study on the XAI question bank \cite[p.6]{Liao2020}:
\begin{quote} [explanations by system rationale] are essentially `this is how I do it, take it or leave it'. But doctors don't like this approach...Thinking that [AI is] giving treatment recommendations is the wrong place to start, because doctors know how to do it. It's everything that happens around that decision they need help with... more discussions about the output, rather than how you get there. \end{quote}
Although the XAI question bank allows for the design of more human-centred XAI, it is still very algorithmic- and data-centred, as the authors themselves state and as we will discuss below. Even if it is known how a DSS arrived at an output or recommendation, operators might be more interested in factors regarding the broader socio-technical context outside of the model \cite{Zednik2019, Ehsan2021, Ghassemi2021, Ehsan2024}, which could relate to questions like, ``What is missing from the dataset?'', ``Are there any other overlooked alternative solutions?'', or ``How to integrate this recommendation into the decision-process?''. 

With our proposed taxonomy, we want to stimulate ``more discussions about the output'' \cite[study participant in][]{Liao2020} by identifying and categorising relevant questions. 
The XAI question bank will prove useful as it contains questions of operators and maps them to XAI techniques. 
As we will show, we utilise the mapping from questions to XAI techniques and reverse the process to repurpose explanations to formulate possible questions, e.g., ``Is it relevant to focus on factor $x$?''.
As such, in line with actionability, we consider ``explanations [informing questions] as a means to help a data subject [and decision-maker] \textit{act} rather than merely understand'' \cite[p.843]{Wachter2017}, where `act' in this case is deliberate decision-making through reflection.

\subsection{Increasing Cognitive Engagement}
\label{alternative-hci} 

To mitigate overreliance, scholars explore different interactions with machine recommendations and explanations in addition to XAI that attempt to cognitively engage the decision-maker in the decision-making process. 
We will briefly mention four relevant approaches: 1) promoting cognitive engagement through interventions, 2) supporting the decision-maker to make their own decision, 3) formulating explanations as questions, and 4) proposals to stimulate reflection.

First, cognitive interventions are based on the dual-process model of reasoning, and aim to interrupt the decision-maker's habitual thinking (system 1) in order to encourage analytical thinking (system 2) \cite{Lambe2016}. Cognitive interventions can take the form of checklists, instructions for analytical thinking, or reflection, for example \cite{Lambe2016, Prakash2019}. In the context of human-machine interaction, one study found that such interventions, i.e., presenting explanations delayed, on-demand, or only after the operator made a decision, can reduce, yet not eliminate, overreliance on DSS \cite{Bucinca2021}. 

Second, instead of presenting a recommendation and justifying it with an explanation that the decision-maker must accept or reject, a so-called hypothesis-driven recommender shows evidence for and against a decision \cite{Miller2023}. In this way, the system supports the decision-maker's reasoning process and allows them to make an informed decision themselves. A study found that while presenting evidence for and against did little to improve diagnostic accuracy, physicians valued the reflective aspect and had more confidence in their final decisions \cite{Cabitza2023}.

Third, one study found that explanations formulated as questions improve human judgement of the logical validity of the information provided \cite{Danry2023}. Causal explanations were presented both as statements and questions. For example, the statement: ``The patient has the flu because of a headache'', is formulated as a question: ``If the patient has a headache, does it follow that they have the flu?''.
Similar to the hypothesis-driven recommender, questions help the decision-maker to think for themselves \cite{Ryan2000}, which leads to more cognitive engagement \cite{Gajos2022}. 

Fourth, it has been suggested that DSS should take on roles other than that of a recommender, such as a devil's advocate raising objections and challenging the decision-maker \cite{Chiang2024, Ma2024, Sarkar2024}. Insights from decision-making in human groups show that a devil's advocate has positive effects on decision performance \cite{Schwenk1984, Schweiger1989}. Similarly, others argue that DSS should support deliberation and facilitate reflection \cite{Cornelissen2022, Haselager2023, Schmidt2023, Zhang2023, Ma2025}. 

In view of the above, we understand questions as cognitive interventions (1), which facilitate own reasoning and decision-making (2\&3). With our taxonomy, we therefore want to both build on and expand these four approaches by offering more potential questions compared to (3) and supporting  the design of suggested interactions that promote reflection (4).

\subsection{Reflection and Responsible Decision-Making}
\label{reflection}

Amongst various forms of increasing cognitive engagement, such as checklists or feedback, stimulating reflection proved to be the most effective \cite{Lambe2016, Prakash2019}.
While reflection can occur before, during or after an action \cite{Rogers2001}, we focus on reflection during decision-making, i.e., reflection-in-action \cite{Schon1983}. As mentioned earlier, reflection can promote reasoning, critical thinking and problem-solving, and thus improve decision-making.

In areas where decisions have far-reaching consequences, such as law or healthcare, decision-makers have great professional responsibility. Physicians, for example, are committed to the well-being of patients. This professional responsibility includes an epistemological responsibility that requires decision-makers to gather and evaluate relevant information, and to know the reasons for a particular course of action \cite{vanBaalen2015}. Being able to provide reasons for (past) actions allows for backward-looking responsibility or accountability. More importantly, knowing the reasons enables one to determine and direct future actions to achieve a desired state, and thus to take forward-looking responsibility \cite{Loughran1996}. 

Reflection-in-action can therefore increase epistemological and thus professional responsibility, both in a backward-looking and, more importantly, in a forward-looking sense. Ideally, this forward-looking responsibility leads to fairer and more desirable decision outcomes for all parties involved \cite{Ueda2024}. 
Reflection-in-action could, for example, prevent a physician from jumping to conclusions after seeing seemingly similar patient cases throughout the day, which could lead to tunnel vision \cite{Mamede2023}. Reflection makes it possible to distinguish between relevant and irrelevant information, and, to quote \citet[p.17]{Dewey1933} again, ``converts action that is merely appetitive, blind and impulsive into intelligent action''.

\subsection{Types of Questions}
\label{socratic-method}

A common technique to facilitate reflection is to ask questions \cite{Osmond2005}. Questions encourage independent thinking, which can increase confidence in one's own decision-making, i.e., decision self-efficacy. This in turn can affect the motivation and well-being of the decision-maker \citep{Ryan2000}. Findings from educational research suggest that motivation functions as an antecedent to cognitive engagement and also increases it \cite{Blumenfeld2006, Rotgans2011, Singh2022}. Questions thus seem to promote reflection, which can increase decision self-efficacy and motivation, which in turn can increase cognitive engagement. 

Given the breadth of possible questions, we want to point out two works from the field of education that categorise questions in different ways and which we also use for our purposes, namely a taxonomy of Socratic questions \cite{Paul2019}, and Bloom's taxonomy for educational objectives \cite{Anderson2001,Krathwohl2002}.

First, Socratic questioning is widely used in education to develop and promote critical thinking skills \cite{Torabizadeh2018, Makhene2019, Dalim2022, Ho2023}. It consists of systematically asking questions that help to clarify concepts, improve understanding, and uncover gaps in knowledge \cite{Paul2019}. Accordingly, Socratic questioning helps to discover one's thoughts, to analyse and evaluate assumptions, information and inferences, and to arrive at judgements through own reasoning. To support teachers in the classroom, a taxonomy of Socratic questions has been developed, defining eight types of questions that probe different parts of thinking, such as assumptions, reasons, or implications \cite{Paul2019}.

Second, Bloom's taxonomy categorises educational learning objectives hierarchically into six levels of cognitive processes, which are: \textit{remembering, understanding, applying, analysing, evaluating, and creating}. The last three levels, analysing, evaluating, and creating, are also components of critical thinking \cite{Paul2019}. This categorisation will be helpful when formulating questions that address different cognitive processes.

In the context of human-machine interaction, the Socratic questioning method is slowly gaining traction.
Based on the taxonomy of Socratic questions, a dataset of 110k questions that could be used in counselling and coaching is proposed, including a generation method of prompt-tuning a language model \cite{Ang2023}. The resulting questions fall into one of the six categories defined by \citet{Paul2019}: 1) clarification (``What do you mean by...?''), 2) probing assumptions (``Why do you assume...?''), 3) probing reasons (``How do you know that...?''), 4) probing implications (``If..., what is likely to happen?''), 5) probing alternatives (``What else should we consider?''), and 6) others that do not conform to the Socratic categories \cite{Ang2023}. In addition, a virtual assistant that fosters moral enhancement through Socratic questions is proposed \cite{Lara2020}, as well as chatbots based on language models that utilise the Socratic method for thought-provoking tutoring \cite{Favero2024, Liu2024b}.

However, there is currently no systematic approach to the use of (Socratic) questions to promote critical reflection in decision-making, particularly in conjunction with decision-support systems. Given the link between critical reflection and reasoning and the associated positive effect on decision-making, we see potential for the use of questions to promote decision self-efficacy and cognitive engagement.
We therefore aim to address this gap by adapting the taxonomy of Socratic questions \cite{Paul2010, Paul2019} to the context of machine-assisted decision-making. In doing so, we also aim for a higher level of granularity as compared to \citet{Ang2023}, who utilise parts of the Socratic questions taxonomy in unchanged form to generate counselling and coaching questions.

\section{Methodology}

Our taxonomy is based on the taxonomy of Socratic questions \cite{Paul2019}, the XAI question bank \cite{Liao2022}, and a revision of Bloom's taxonomy \cite{Anderson2001} discussed in the previous section. 
To construct our taxonomy we proceeded in two steps: 1) identifying elements that questions can relate to, and 2) identifying information for formulating and enriching questions. 
In order to identify elements (1), we synthesised the taxonomy of Socratic questions with the XAI question bank through a procedure similar to affinity diagramming.
In order to help formulate and enrich questions (2), we reversed the mapping of the XAI question bank to repurpose information derived from explanations. In addition, we used the cognitive processes identified in Bloom's taxonomy to determine the scope of questions. 

\subsection{Procedure}

\subsubsection{Phase 1: Identifying Elements for Reflection}

First, we identified relevant aspects of machine-assisted decision-making to which potential questions should relate, i.e., \textit{what} to reflect on. In a deductive (top-down) approach, the categories from the taxonomy of Socratic questions served as a starting point. 
As Elder and \citet{Paul2010} note:
\begin{quote}
    ``Whenever we think we think for a \textit{purpose} within a \textit{point of view} based on \textit{assumptions} leading to \textit{implications and consequences}. We use \textit{data, facts and experiences} to make \textit{inferences} and judgements based on \textit{concepts} and theories to answer a \textit{question} or solve a problem.''
\end{quote}
Based on this structure of thinking, the authors categorise questions into elements of thought or reasoning as follows:
\begin{itemize}
    \item \textbf{Purpose}: Questions about the agenda, goal or objective.
    \item \textbf{Question at Issue}: Questions about the problem or issue that gave rise to the question.
    \item \textbf{Information}: Questions about background information, such as data, facts, observations or experiences.
    \item \textbf{Interpretation and Inferences}: Questions about how meaning was derived and conclusions drawn.
    \item \textbf{Concepts}: Questions about underlying theories, definitions, or models that define a thought.
    \item \textbf{Assumptions}: Questions about taken for granted presuppositions of a thought.
    \item \textbf{Implications and Consequences}: Questions about implications.
    \item \textbf{Point of View}: Questions about the frame of reference or perspective of a thought.
\end{itemize}
While these elements of thought, including the provided example questions (Table~\ref{table:socratic-questions}), are helpful in general, they cannot be directly applied to the domain of decision-making (with DSSs). This is because the taxonomy of Socratic Questions is primarily intended for teachers and students in the classroom. 

We thus took the elements of thought as a starting point for our taxonomy and mapped them to the context of machine-assisted decision-making. For this mapping and to identify relevant aspects of machine-assisted decision-making, we used the XAI question bank. The question bank contains more than 50 common questions that operators have towards a DSS, categorised as follows \cite{Liao2022}:
\begin{itemize}
    \item \textbf{How} (global model-wide): Questions about the general logic of the model.
    \item \textbf{Why} (a given prediction): Questions about the reasons for a prediction.
    \item \textbf{Why Not} (a different prediction): Questions about the difference to an expected outcome.
    \item \textbf{How to be That} (a different prediction): Questions about how to change an instance to get a different outcome.
    \item \textbf{How to still be This} (the current prediction): Questions about possible changes to still get the same outcome.
    \item \textbf{What if}: Questions about changed outcome based on changed input.
    \item \textbf{Performance}: Questions about the performance of the DSS.
    \item \textbf{Data}: Questions about training data.
    \item \textbf{Output}: Questions about how to apply or use the DSS output.
\end{itemize}
Overall, questions from the XAI question bank relate to the logic of the model, its performance, the reasons for a prediction, possible changes, the data, and the output.

\begin{figure}[t]
    \centering
    \includegraphics[width=0.7\linewidth]{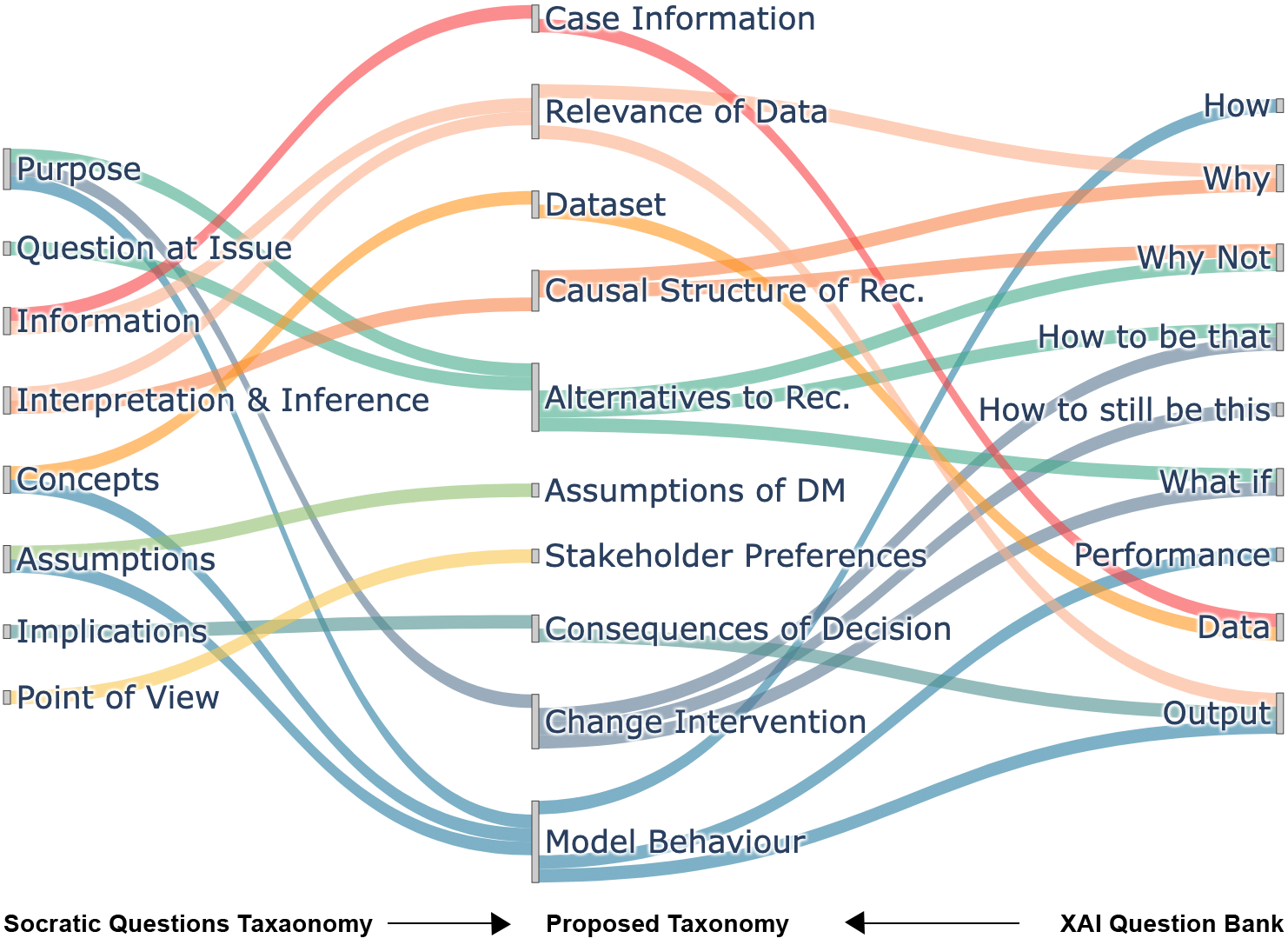}
    \caption{Our elements for reflection in machine-assisted decision-making in the middle column (see section~\ref{taxonomy}) are based on the taxonomy for Socratic questions \cite{Paul2019}, in the left column, and the XAI question bank \cite{Liao2022}, in the right column.}
    \label{fig:mapping}
    \Description{A sankey plot showing the connections between the elements of our taxonomy and the taxonomy for Socratic questions and the XAI question bank.}
\end{figure}

This leaves us with the categorisation of questions based on thought structure on the one hand, and the components of machine-assisted decision-making represented by clusters of common questions in the context of explainable AI on the other (Figure~\ref{fig:mapping}). The element \textit{Information}, for example, can be mapped, as is also mentioned in the definition, to \textit{Data} the DSS is trained on, as well as (input) data of a specific case. Through affinity diagramming we arrived at the following 10 elements, which are shown in the middle column of Figure~\ref{fig:mapping}: \textit{case information}, \textit{relevance of data}, \textit{dataset}, \textit{causal structure of recommendation}, \textit{alternatives to recommendation}, \textit{assumptions of decision-maker}, \textit{stakeholder preferences}, \textit{consequences of decision}, \textit{change intervention}, and \textit{model behaviour}. 

\subsubsection{Phase 2: Formulating Questions}

After identifying \textit{what} to question, the next step was to specify \textit{how} to formulate and enrich specific questions. For this, we again used the XAI question bank, in particular the mapping of common questions to XAI techniques.
Instead of starting with questions from operators in order to arrive at suitable explanations, however, we reversed the process. We moved from explanations to questions, in order to extract information using XAI techniques, which then flows into possible questions.
On the one hand, this means that (some) explanations can be (re-)formulated as questions, e.g., ``Does outcome $Y$ follow from information $x$?'' \cite{Danry2023}. On the other hand, questions can relate to information extracted from explanations, e.g., ``Is information $x$ a relevant factor?''. 

Finally, to capture the scope of questions, we used, albeit only marginally, a revised version of Bloom's taxonomy. In Bloom's taxonomy, questions are categorised into levels of cognitive processes to support the formulation of learning objectives. 
Three of these categories or cognitive abilities are components of critical thinking \cite{Paul2019}, namely, analysing, evaluating, and creating, where \cite{Anderson2001}:
\begin{itemize}
    \item \textbf{Analysing} means to break down information into parts and determining how these parts are connected, and identifying which parts are relevant or irrelevant. Verbs that relate to this cognitive process are: \textit{compare, criticize, differentiate, discriminate, deconstruct, inspect}. For example, ``What evidence is there for $Y$?''. 
    \item \textbf{Evaluating} means to judge information and its appropriateness, and to detect inconsistencies. Verbs that relate to this cognitive process are: \textit{assess, support, defend, validate, evaluate}. For example, ``How to justify $Y$?''
    \item \textbf{Creating} means compiling information in a new way, or considering alternative hypotheses. Verbs that relate to this cognitive process are: \textit{create, develop, formulate}. For example, ``What could be done to minimise $Y$?''
\end{itemize}
Based on this cognitive distinction, we suggest that questions should address these higher-level cognitive processes in order to effectively stimulate critical reflection.

\subsection{Expert Feedback}

While one author created the initial taxonomy, we discussed the categories with their overlaps and boundaries among the authors throughout the process. Besides, we created the taxonomy in an iterative way. Once we had a first version, we sent 10 initial categories including some sample questions to one orthopaedic surgeon and one senior researcher in orthopaedics, who both work at different hospitals and are part of an ongoing collaboration. To get some early feedback, we asked them if they saw any missing question types. They confirmed the completeness of the derived question types and mentioned that it seems we ``covered all the aspects'' and that they ``don’t see any obvious questions missing''. We then refined the categories, without adding or removing any, expanded details, added descriptions, and added sample questions to a separate table (Table~\ref{table:example-questions-health}). We presented the final taxonomy again to the same surgeon and researcher (section~\ref{evaluation}).

\section{Taxonomy of Questions}
\label{taxonomy}

Our taxonomy consists of 10 question types or elements for reflection (Table~\ref{table:taxonomy}). In order to demonstrate how to apply our taxonomy, we provide some sample questions from the medical (Table~\ref{table:example-questions-health}) and educational domain (Table~\ref{table:example-questions-education}). A description of the individual question type follows, focussing on the medical field, where a physician uses a DSS to make a decision about a patient's treatment.

\begin{table*}
\caption{The taxonomy we propose helps to systematically formulate questions that stimulate critical reflection in machine-assisted decision-making. From a taxonomy of Socratic questions we take the elements of thought \cite{Paul2019}, i.e., the \textit{What}, indicated in \textit{italic}, and adapt them to the context of machine-assisted decision-making. To identify information for enriching questions, i.e., the \textit{How}, we use the XAI question bank \cite{Liao2020,Liao2021,Liao2022} and reverse the mapping between questions and explanations. Identifiers (IDs) marked with an asterisk (*) indicate questions that address the level of creating, according to Bloom's taxonomy \cite{Krathwohl2002}. For sample questions, we refer to Table~\ref{table:example-questions-health} and Table~\ref{table:example-questions-education} in the Appendix.}
    \label{table:taxonomy}
\begin{tabulary}{\linewidth}{@{}LLLL@{}}
\toprule
\textbf{ID} &
  \textbf{Element for Reflection} (What?)&
  \textbf{Description} &
  \textbf{Useful Information} (How?)
  \\
\midrule
Q1 &
Case Information \newline(\textit{Information}) &
  Questions that help further assess, inspect, and contextualise data points to ensure its quality, reliability, and completeness. &
  Input Data (e.g., tabular patient data), Datasheets \cite{Gebru2021}
  \\
Q2 &
Relevance of Data \newline(\textit{Information, Interpretation \& Inference})&
  Questions that help evaluate whether the data adequately supports the hypothesis / recommendation. &
  Feature Contribution (e.g., SHAP \cite{Lundberg2017}, LIME \cite{Ribeiro2016}) 
  \\
Q3 &
Dataset \newline(\textit{Concepts}) &
  Questions that help analyse and assess the data and to ensure it adequately represents the phenomenon (by highlighting limitations or characteristics of the dataset). & 
  Training Data, Datasheets \cite{Gebru2021}, FactSheets \cite{Arnold2019}
  \\
Q4 &
Causal Structure of Recommendation \newline(\textit{Interpretation \& Inference}) &
   Questions that help evaluate whether the outcome follows from data to ensure that the causal structure of the model / recommendation is sound.
   &
  Feature Contribution \cite{Lundberg2017, Ribeiro2016}, Counterfactuals \cite{Karimi2021,Poyiadzi2020}
  \\
Q5* &
Alternatives to Recommendation \newline(\textit{Question, Purpose})&
  Questions that help to consider other possibilities ensuring that the larger solution space is considered. &
  Contextual information (e.g., differential diagnosis or patient preferences (Q7)), Partial Dependence Plot, Counterfactuals
  \\
Q6 &
Assumptions and Expectations of Decision-Maker \newline(\textit{Assumptions})&
  Questions that elicit taken-for-granted assumptions to ensure that the decision-maker is aware of their reasons and possible cognitive biases. &
  General questions, Cognitive biases, Confidence of decision-maker
  \\
Q7* &
Stakeholder Preferences \newline(\textit{Point of View}) &
  Questions that help the decision-maker enquire and take into account the preferences or needs of the people concerned (e.g., patient), in order to ensure a better or broader understanding of the problem at hand. &
  General questions, Preference elicitation, Values
  \\
Q8* &
Consequences of Decision \newline(\textit{Implications}) &
  Questions that help elicit anticipation and forward-looking responsibility to ensure that (unintended) consequences and trade-offs are considered and mitigated. & 
  Limitations of Model, FactSheets \cite{Arnold2019}, Model Cards \cite{Mitchell2019a}
  \\
Q9* &
Change Intervention \newline(\textit{Purpose}) &
  Questions that support the exploration of feasible (smaller) interventions to make a desired outcome more likely. &
  Feature Perturbation, Counterfactuals
  \\
Q10 &
Model Behaviour \newline(\textit{Assumptions, Purpose, Concepts}) &
  Questions that help to assess and evaluate the assumptions, rules and thresholds built into the model. In particular, areas of decision limits, i.e. until when does the result remain the same and when does it change. &
  Decision Boundaries, Feature Perturbation, Counterfactuals, Model Cards \cite{Mitchell2019a}, Global Feature Importance \cite{Lundberg2020}
\\
\bottomrule
\end{tabulary}
\end{table*}

\textit{Q1}. Questions can relate to the \textit{case information} provided, which represents information in the form of data.
Previous operations on the spine, for example, might be indicated by integers. This value, however, does not indicate how long ago the operation was performed and at which location of the spine it was carried out, both of which can influence the effectiveness of further operations.
Questions can help the decision-maker to inspect and contextualise the data provided to give it semantic depth.
To ensure data quality, questions can also relate to the reliability of the data provided to rule out inconsistencies, such as measurement errors, and to ensure that no relevant data is missing.
As such, Q1 helps the decision-maker get a more holistic picture of the case \cite{vanBaalen2015}.
  
\textit{Q2}. Questions can relate to the \textit{relevance of data}. 
In doing so, questions can encourage the decision-maker to contextualise the information, e.g., symptom, in a specific case, which can lead to validating the machine recommendation by other means, e.g., by conducting additional tests. 
So while XAI techniques, such as feature importance, can show which features have contributed to an outcome, domain knowledge is necessary to evaluate whether these data points are important in a particular case. It is possible to use the information extracted with XAI to generate an evaluative question, like ``How relevant is data $x$?''. 
The question of the relevance of the most important data point(s) also raises the question of whether there are other factors that could be overlooked in the decision.
As different XAI techniques can result in different rankings of the contributing effects of the individual features \cite{Saarela2021}, it is important for decision-makers not to take the information obtained through XAI at face value. The discrepancies between two opposing feature ranks from two different XAI methods could also lead to a question like ``Is $x$ or $z$ more relevant?''. 

\textit{Q3}. Questions can relate to the \textit{dataset} on which the model was built. 
Cardiovascular disease, for example, has different symptoms in men and women, yet most of the data for training DSSs stem from male patients, which is why DSSs often misdiagnose women \cite{Li2023a}. 
Certain populations or phenomena are often overrepresented in data, which leads to a skewed data distribution \cite{Ueda2024}. Accordingly, a reflective question could be raised if there are large differences between the case information and the dataset. Besides, data is often only a partial representation of the phenomena under investigation \citep{Broussard2019,Binns2018}, as it is impossible to quantify every aspect of the complexity of social life \cite{Birhane2021a}. 
Questions can help to consider limitations of a dataset and take into account information that is not yet included and is (now) known to contribute to the outcome.
So compared to Q2, which is more concerned with the question of what information is relevant for a single recommendation (local), Q3 asks whether the dataset contains relevant information (global). 

\textit{Q4}. Questions can relate to the (apparent) \textit{causal structure} between provided case information and resulted machine recommendation. 
Again, based on feature contribution it is possible to ask whether and how the machine recommendation follows from data point $x$, for instance ``Does diagnosis $Y$ follow from symptom $x$?''. So although this question type merely reformulates a causal explanation \cite{Danry2023}, the aim is to help the decision-maker to scrutinise the information.

\textit{Q5}. Questions can help identify and consider \textit{alternatives to the recommendation}. This calls into question the appropriateness of a recommendation.
It might be possible to treat the same condition in different ways, where the feasibility of each option depends on various constraints like available resources or the patient's preferences (Q7).
For example, one purpose might be full recovery, for which a DSS might recommend surgery that involves a long hospital stay; another purpose might be to promote the patient's well-being, who might prefer to spend as little time as possible in hospital, so that the more appropriate treatment might be a non-surgical intervention. 
At the same time, as different disease have common symptoms, the physician wants to ensure that another possible diagnosis is not overlooked, e.g., through a differential diagnosis \cite{Mamede2023}.

\textit{Q6}. Questions can relate to the \textit{assumptions of the decision-maker}. On the one hand, these questions can prevent the decision-maker from jumping to conclusions by helping to uncover tacit knowledge, taken for granted assumptions, and cognitive biases and re-evaluate them case by case. 
On the other hand, questions can help to increase the confidence and self-efficacy of decision-makers as they allow them to make their own decision. 

\textit{Q7}. Questions can relate to \textit{stakeholder preferences}.
If the decision-maker is not the decision-subject, as is the case with physicians and patients, it is important to consider the different perspectives.
Ideally, enquiring about the preferences of stakeholders is part of the decision-maker's routine practice, yet these questions can function as a form of safeguard. If the decision-maker is the decision-subject, questions can help elicit preferences and values of the decision-maker themselves.

\textit{Q8}. Questions can help anticipate (unintended) \textit{consequences of a decision}. 
While the decision-maker has ideally weighed up the implications, these questions can promote forward-looking responsibility of the decision-maker. Furthermore, with insights from the dataset (Q3) attention could be drawn to certain limitations of a DSS. Compared to Q3, Q8 focuses on assessing the impact of a recommendation, weighing trade-offs, and supporting the decision-maker in considering appropriate mitigation strategies.

\textit{Q9}. Questions can relate to \textit{change interventions} in order to further investigate the feasibility of desired alternatives.
For such questions, hypothetical scenarios can be created by feature perturbation and counterfactual explanations. Unlike counterfactual explanations, which provide information about what the outcome would have been if the data had been like this, these questions are about whether it is possible to change certain data points to make a particular outcome more likely.
For example, a question might help the physician to consider prior and smaller interventions such as  smoking cessation, which could increase the expected effectiveness of a desired treatment such as surgery.
Although these questions ideally lead to more actionable insights, they might imply unfeasible or unattainable changes and actions, e.g., changing the age of a patient. These hypothetical scenarios can nevertheless provide insights into inbuilt associations and decision boundaries of the DSS.

\textit{Q10}. Questions can relate to the \textit{model behaviour}.
These questions help to clarify the assumptions built into the DSS, which may be based on the data set (Q3) or other assumptions made during model development. As such, these questions can also provide the decision-maker with a better understanding of the workings and (apparent) causal structure built into the model (Q4). For example, if a DSS makes a decision based on age with a threshold of 50 years, the decision-maker may want to re-evaluate the decision, if the patient is close to that threshold, such as 48 years old.

\section{Evaluation}
\label{evaluation}

To formatively evaluate our question taxonomy, we presented a first version with sample questions and the final version in two phases to the same orthopaedic surgeon and orthopaedic researcher. 
In individual online meetings of 30 minutes each, which were not recorded or transcribed, we asked them whether the taxonomy and the derived questions could help practitioners, whether the considerations associated with these questions already correspond to common practice, whether certain questions might be more relevant than others, and whether they could provide example questions or example cases. 

Overall, both recognised the relevance of our taxonomy and the use of questions. They were confident that questions could help to re-evaluate initial assumptions of the physician, and counteract habitual priming, e.g., seeing similar yet different patient cases throughout the day. Importantly, as the surgeon noted, this priming also happens through the information presented by a DSS, which questions could help to mitigate. Moreover, they mentioned that the questions derived from the taxonomy could help to treat different patients more objectively by reducing discrepancies between cases, such as different patients describing their pain differently, i.e., pain catastrophising. 
As one remarked, given that many question types from the taxonomy are derived from data, it is difficult to ask about data points that are not contained in the dataset (Q3).\footnote{In this case, general questions like ``Were all relevant information considered?'' or ``Did you consider factor \textit{x} which is known to influence the outcome, but is not part of the data set'' could help.} To derive relevant questions, they suggested to talk to experts and ask for their reasoning process, as the usefulness of specific questions, as they noted, depend on the specific use case. Accordingly, they also pointed out that questions should help and not irritate.

\subsection{Transferability of Our Taxonomy}

To assess the applicability of our taxonomy in another domain, we discussed the finalised version with two postdoctoral researchers in the field of education and technology, and one doctoral researcher who is also a primary school teacher. These individual and informal conversations, each lasting about 30 minutes, were neither recorded nor transcribed. The aim of these conversations was to clarify whether the taxonomy is relevant for the educational domain, and to familiarise ourselves with some exemplary use cases and questions.

The three researchers shared two examples of so-called learning analytics dashboards (LAD), which educators can use to plan lessons, define learning objectives, track the learning progress of students, create exercises accordingly, and grade students.\footnote{\url{https://snappet.org/} and \url{https://www.gynzy.com/en}.} These LADs also recommend exercises and learning objectives based on student performance that teachers can use.
In view of this, one researcher commented, ``Teachers feel the need to be more critical, but do not know how'' as they wonder whether they use the LAD correctly. Accordingly, the researchers noted that questions could contribute to the interpretation of the student's progress shown in the LAD, and help consider factors that go beyond data points, as these systems ``do not know everything''. All three researchers therefore confirmed the expected relevance and applicability of our taxonomy for stimulating teachers' critical reflection in making decisions about students and their futures using recommendations of LADs. 

After the three feedback sessions, we created sample questions for the educational domain (Table~\ref{table:example-questions-education}). We sent these to the three researchers and two of them added some questions and comments that helped us refine our derived questions.
We then presented the sample questions, including our taxonomy, in a lab session attended by 12 people, from doctoral researchers to full professors. The research group, which explores the impacts of technologies on learning and teaching practices, recognised the usefulness of our taxonomy. As one participant commented, the sample questions we created are very close to the questions teachers already ask themselves. Nevertheless, the same researcher continued, our taxonomy could be informative to see which questions are covered and which are left out. In addition, another researcher noted that it is very promising that our taxonomy aligns with the questions that teachers ask in relation to algorithmic learning goal recommendations. Furthermore, one attendee noted that students doing the exercises in a LAD could also benefit from questions for critical reflection, such as considering the benefit of doing an exercise that might go against (short-term) preferences.

In comparison to the medical sample questions (Table~\ref{table:example-questions-health}), the wording of the educational sample questions differs. The structure and elements of critical reflection however remain the same for both. We therefore maintain that our taxonomy is transferable and can be applied to other decision-making domains without adaptation as it builds on general concepts of critical thinking.

\subsection{Effectiveness of Questions}

Another type of evaluation, beyond the scope of this paper, is to assess the effectiveness of the questions created based on the taxonomy in decision-making, and the extent to which they promote critical reflection and cognitive engagement of the decision-maker.
Existing studies, primarily from the educational domain, point to the positive effects of Socratic questioning strategies on moral reasoning in nursing \cite{Torabizadeh2018} and critical thinking skills during (healthcare) education of students \cite{Ishtiaq2024, Jameel2025, Yang2005, Makhene2019, Ho2023}. These studies demonstrate the validity and robustness of the Socratic questioning approach on which our proposed taxonomy is based, which makes the prospects for our taxonomy promising. 

Nevertheless, assessing the effect of questions on decision-making remains an open challenge. Current evaluation methods for decision-making and overreliance on DSS typically only focus on the outcome of a decision, i.e., decision accuracy \cite{Guo2024,Klingbeil2024}. To measure reflection and cognitive engagement, however, it is important to consider the entire decision-making process, including the reasoning process, whether alternatives were considered, the decision-maker's confidence in the final decision, and satisfaction with the outcome. Existing methods for measuring cognitive engagement are associated with considerable effort and technical barriers, such as electroencephalography (EEG) \cite{Hassib2017, Berka2007}, or the measurement of pupil dilation \cite{VanDerWel2018}.
Consequently, an accessible evaluation framework would need to be developed in future work, e.g., a self-report scale that takes the entire decision-making process into account. For this, inspiration can be drawn from existing scales for factors associated with cognitive engagement, such as self-efficacy, motivation, and confidence \cite{Lee2025, Hart2014, Greene2015}, as well as from methods in the educational domain for assessing student's cognitive engagement with technology \cite{Vongkulluksn2022}.

\section{Discussion}
\label{discussion}

We have presented a taxonomy of question types to support the systematic creation of questions that stimulate critical reflection in machine-assisted decision-making.
As Figure~\ref{fig:mapping} shows, the questions identified in the XAI question bank for human-centred XAI are very model- and data-centric. As a result, our taxonomy is more data-oriented. Nevertheless, our taxonomy can help to take into account human factors such as assumptions or cognitive biases of the decision-maker and stakeholder preferences as well as consequences of a decision to overcome the current algorithmic focus of XAI (section~\ref{human-xai}). In doing so, questions can stimulate discussion about the machine recommendation and its integration into the broader decision-making process.

\subsection{Different Stakeholders}

Our taxonomy is primarily targeted at decision-makers who operate DSS to make informed decisions.
Accordingly, the aim of our taxonomy is not to formulate questions for the decision-maker to assess the performance of the DSS as this would unfairly shift the responsibility from developers and regulators to operators. The deployed models must be evaluated and tested beforehand, which requires a shift from individual human oversight to institutional oversight \cite{Green2022}.
Considering this, other stakeholders could also benefit from critical reflection and thus our taxonomy. 
\textit{Developers} constructing a model, for example, might ask themselves what assumptions they are making and whether the dataset used is suitable for solving a given problem.
\textit{Auditors} assessing compliance could benefit from our taxonomy when defining objectives for an audit of the model or dataset \cite{Raji2020, Birhane2024b}. 
Moreover, the taxonomy could be used as a form of supporting AI literacy to help people remain more critical of machine-generated information, especially in view of misinformation \cite{Zavolokina2024}. \textit{Patients} could be supported in identifying (treatment) preferences through reflection, thereby asking relevant questions to their healthcare providers (operating DSS), or being critical while interacting with a medical chatbot themselves. \textit{Students} could be taught to question the relevance of retrieved information and could be helped to remain cognitively engaged in their technology-aided learning.
Our proposed taxonomy could therefore benefit various stakeholders involved in the design, development, testing, and use of DSS or machine learning systems.

\subsection{Creation of Questions}

To facilitate the formulation of questions, we suggested that questions should address the levels of critical thinking according to Bloom's taxonomy, i.e., evaluating, analysing, and creating. 
As such, it might be that questions cannot be answered directly and instead lead to further questions \cite{Paul2019}. To motivate further inquiry and engagement of the decision-maker, a brief explanation of why a question should be entertained could be given, as mentioned by the orthopaedic surgeon we spoke to. 

It must be considered that questions can be leading and possibly steer the decision-maker towards an undesirable decision. Hence, we advise against the use of large language models, as the generated questions could be irrelevant and unreliable \cite{Weidinger2022}. Instead, researchers or designers ideally work together with experts to create useful questions.
When creating the sample questions for the educational domain we realised that it is difficult to do so without having much knowledge about the respective use case, access to data, or the functionality of the DSS or learning analytics dashboard (LAD). Nevertheless, uncovering these blind spots might enable further inquiry into how the system works, and open up a dialogue between developers or service providers and domain experts or affected organisations, such as hospitals or schools. 

Furthermore, certain questions might fall into multiple categories of our taxonomy, as different question types can overlap or build on other categories. The causal structure of a recommendation (Q4), for example, can entail questions about the data (Q1\&Q3). We believe, however, that this is not a problem as it shows how different elements are connected to each other and it could rather help to formulate questions more precisely so that they can be delineated.

\subsection{Presentation of Questions}

To present the questions to decision-makers, one of the educational researchers we spoke to suggested that a poster of our taxonomy with sample questions could be displayed in the staff room to sensitise teachers to the different elements of the machine-assisted decision-making process. 
Alongside an analogue form and building on the above-mentioned proposals of DSS supporting deliberation and reflection (section~\ref{alternative-hci}), our taxonomy could inform the design of a technical implementation. As such, we introduced and envision a \textit{reflection machine} (RM) that functions as an additional system or component of a DSS \cite{Haselager2023, Cornelissen2022, Fischer2025}. Based on input data, the recommendations, the functioning of the DSS derived from XAI, and the use of our taxonomy relevant questions can be created and presented to the decision-maker accordingly.

Supporting the decision-making process with questions might be an unfamiliar interaction. The design of the interface and the integration into existing workflows therefore influence the acceptance and perceived usefulness of a specific prototype \cite{Lund2001}, and thus the effectiveness of questions on cognitive engagement.

\subsection{Cognitive Load}

While future work needs to evaluate the effectiveness of questions, questions could interrupt the workflow and increase the cognitive load of the decision-maker. In the aforementioned study, in which explanations were presented at different times as a form of cognitive intervention, participants reported increased cognitive load, which is why they favoured these interventions less \cite{Bucinca2021}. 
Ultimately, however, the cognitive burden is unavoidable and inherent to critical reflection, as \citet[p.17]{Dewey1933} notes:
\begin{quote} [...] yet thinking need not be reflective. For the person may not be sufficiently critical about the ideas that occur to him. He may jump at a conclusion without weighing the grounds on which it rests; he may forego or unduly shorten the act of hunting, inquiring; he may take the first `answer' or solution, that comes to him because of mental sloth, torpor, impatience to get something settled. One can think reflectively only when one is willing to endure suspense and to undergo the trouble of searching. \end{quote}
Nevertheless, the right balance between effort and effectiveness must be found. Questions must be well integrated into decision-making processes in order to enable the operator to constructively take the posed question into account. The competences, decision-making style, and need for cognition of the decision-maker play a role here.

\section{Conclusion}

Our taxonomy of 10 question types shifts the emphasis from machine recommendations and explanations to questions and human reflection.
Careful selection and presentation of questions has the potential to increase cognitive engagement, mitigate overreliance on DSS and thus support the effective human oversight required by the European AI Act.

\clearpage
\appendix
\setcounter{table}{0}
\renewcommand{\thetable}{\Alph{section}}
\section{The Socratic Question Taxonomy}

\begin{table*}[h]
    \caption{The following questions are taken from the taxonomy of Socratic Questions \cite{Paul2019}. The authors provide a more comprehensive list of questions.}
    \label{table:socratic-questions}
    \centering
    \begin{tabulary}{\linewidth}{@{}LL@{}}
    \toprule
         \textbf{Element of Thought} &
         \textbf{Sample Questions}
         \\
    \midrule 
         Purpose &
         What are we trying to accomplish? What is the purpose of ...?
         \\
         Question &
         Is this question the best one to focus on? How can we find out?
         \\
         Information &
         On what information are you basing that comment? Are those reasons adequate? 
         \\
         Interpretation \& Inference &
         How did you reach that conclusion? How shall we interpret these data?
         \\
         Concepts &
         What is the main idea you are using in your reasoning? What main theories do we need to consider in figuring out ....?
         \\
         Assumptions &
         What are you taking for granted here? Why do you think the assumption holds here?
         \\
         Implications &
         If we do this, what is likely to happen as a result? What are you implying by that?
         \\
         Viewpoints &
         From what point of view are you looking at this? What would someone who disagrees say?
         \\
    \bottomrule
    \end{tabulary}
\end{table*}

\section{The XAI Question Bank}

\begin{table*}[h]
    \caption{The following questions are taken from the XAI question bank \cite{Liao2021}. The authors provide a more comprehensive list of questions.}
    \label{table:xai-question-bank}
    \centering
    \begin{tabulary}{\linewidth}{@{}LL@{}}
    \toprule
         \textbf{Element} &
         \textbf{Sample Questions}
         \\
    \midrule 
         Data &
         What kind of data was the system trained on? What is the sample size of the training data?
         \\
         Output &
         What kind of output does the system give? How should the output fit in my workflow?
         \\
         Performance &
         How accurate/precise/reliable are the predictions? How often does the system make mistakes? 
         \\
         How (global) &
         How does the system make predictions? What kind of algorithm is used?
         \\
         Why &
         Why/how is this instance given this prediction? What feature(s) of this instance determine the system's prediction of it?
         \\
         Why not &
         Why is this instance NOT predicted to be [a different outcome Q]? Why is this instance predicted [P instead of a different outcome Q]?
         \\
         What if &
         What would the system predict if this instance changes to....? What would the system predict for [a different instance]?
         \\
         How to be that &
         How should this instance change to get a different prediction Q? What is the minimum change required for this instance to get a different prediction Q?
         \\
         How to still be this &
         What is the scope of change permitted for this instance to still get the same prediction? What kind of instance gets the same prediction?
         \\
         Others &
         How/why will the system change/adapt/improve/drift over time? Can I, and if so, how do I, improve the system?
         \\
    \bottomrule
    \end{tabulary}
\end{table*}

\newpage
\section{Our Taxonomy Applied to Healthcare}

\begin{table*}[h]
    \caption{Based on our question taxonomy in Table~\ref{table:taxonomy}, we provide some sample questions that relate to the medical domain, where a physician uses a decision-support system (DSS) in order to find a diagnosis or treatment option for a patient. A collection of general clinical questions serves as input \cite{Ely2000}. Relevant questions will vary depending on the use case, these questions should thus be taken as inspiration.}
    \label{table:example-questions-health}
    \centering
    \begin{tabulary}{\linewidth}{@{}LLL@{}}
    \toprule
         \textbf{ID} & 
         \textbf{Element for Reflection} &
         \textbf{Sample Questions}
         \\
    \midrule
         Q1 & 
         Case Information &
         How does the patient compare against the data distribution? Does the data adequately represent the patient's situation? How do you think the patient understood the self-assessment questionnaire that functions as input to the DSS? When was the indicated surgery performed (previous surgeries decrease the effectiveness of future surgery)? Why did provider $x$ treat the patient this way?
         \\
         Q2 &
         Relevance of Data &
         What are the criteria for diagnosis of condition $y$? Could symptom $x$ be condition y or be a result of condition $y$? What is the likelihood that symptom $x$ is coming from condition $y$? Does feature $x$ [salary] that is part of the data affect the decision?
         \\
         Q3 &
         Dataset &
         Does the dataset adequately represent the phenomena under investigation?  Is the data up-to-date? Is there relevant data missing in the dataset? Did you consider the patient's BMI measure, which the model does not consider, but which influences the result? Where was the data collected? What is the data sample size? Is limitation $z$ of the dataset important to consider (information ideally provided by Datasheet)? Is it important that the data was collected more than 5 years ago?
         \\
         Q4 &
         Causal Structure of Recommendation &
         What is the likelihood that this patient has condition $y$ (given findings $x_1$, $x_2$,..., $x_n$)? Does diagnosis $y$ follow from symptom $x$? Which information supports/contradicts the diagnosis? How good is test $x$ in situation $y$?
         \\
         Q5 &
         Alternatives to Recommendation &
         Are there previous patients with a similar profile who received a different treatment? Are there aspects that might have been overlooked? Is the alternative easy to reject? How do you distinguish between conditions $y_1$, $y_2$? Could this patient have condition $z$ (given findings $x_1$, $x_2$)?
         \\
         Q6 &
         Assumptions and Expectations of Decision-Maker &
         How does the machine recommendation compare to your assumptions? How aggressive/conservative should I be in situation $y$? What are you taking for granted? Are there alternative assumptions you could make? What symptoms are you basing your decision on?
         \\
         Q7 &
         Stakeholder Preferences &
         Does the patient have any preferences that might require a procedure that differs from the recommendation? Does the patient have strong preferences for treatment $y$? Are there certain situational circumstances that prevent the patient from recovering from surgery for several weeks?
         \\
         Q8 &
         Consequences of Decision &
         Are there any unintended consequences of treatment $y$? What are the ethical/legal considerations in situation $y$? What are the administrative considerations in situation y? What were the consequences of treatment $y$ in previous similar patient cases? What are the implications of administering drug $x$?
         \\
         Q9 &
         Change Intervention &
         Is it possible to reduce $x$, so that $y$ becomes more likely? Is it possible to change the patient's expectations to increase the likelihood of effective surgery?
         \\
         Q10 &
         Model Behaviour &
         Would you suggest the same treatment if the patient were 5 years older? Did you consider that the model makes errors in 1 out of 10 cases?
         \\
    \bottomrule
    \end{tabulary}
\end{table*}

\newpage
\section{Our Taxonomy Applied to Education}

\begin{table}[h]
    \caption{Based on our question taxonomy in Table~\ref{table:taxonomy}, we provide some sample questions that relate to an educational use case where teachers use learning analytics dashboards to plan lessons, track student progress, set learning goals, and create exercises. The following questions were created with the help of three researchers in education and technology, one of whom is a teacher. Relevant questions will vary depending on use case, these questions should thus be taken as inspiration.}
    \label{table:example-questions-education}
    \centering
    \begin{tabulary}{\linewidth}{@{}LLL@{}}
    \toprule
         \textbf{ID} & 
         \textbf{Element for Reflection} &
         \textbf{Sample Questions}
         \\
    \midrule
         Q1 & 
         Case Information &
         Does the data adequately represent the student’s progress? Are there other factors outside the LAD that affect the student’s progress?
         \\
         Q2 &
         Relevance of Data &
         Does the failed exercise represent the student's overall performance? Is the recommended exercise relevant to the learning goal? How important is the result of the failed / passed exercise for the overall progress?
         \\
         Q3 &
         Dataset &
         What (student) data was the model trained on? Which metrics are included in the dataset?
         \\
         Q4 &
         Causal Structure of Recommendation &
         Does the recommended exercise contribute to the learning goal? How effective is the recommended exercise for the learning goal?
         \\
         Q5 &
         Alternatives to Recommendation &
         Are there other suitable exercises that the students can do without their devices (what is pedagogically needed for my class)? What other exercise can be done to achieve the learning goal? Should the student work on a learning goal based on their own progress or based on the level of the whole class? 
         \\
         Q6 &
         Assumptions and Expectations of Decision-Maker &
         What are my assumptions about the student? Are my assumptions about the student adequate? 
         \\
         Q7 &
         Stakeholder Preferences &
         Does the student have any special needs? Can the student benefit from other exercises / help? Are there any (extraordinary) circumstances of the student that should be considered?
         \\
         Q8 &
         Consequences of Decision &
         Does the recommended exercise contribute to the learning goal? What does it do to the emotional well-being of my student if they have to repeat exercises multiple times based on LAD recommendations? What are the downsides of this exercise? What other learning objectives are being used in this exercise that the student could be struggling with?
         \\
         Q9 &
         Change Intervention &
         Is it possible to increase the student’s motivation to make the learning goal more likely?         
         \\
         Q10 &
         Model Behaviour &
         Does the LAD contain learning theory / methodology X / adhere to our school-wide policy / goal? Is the model's assumption that the student will not achieve the learning goal if fail two exercises appropriate? 
         \\
    \bottomrule
    \end{tabulary}
\end{table}

\begin{acks}
We thank Sebastiaan van de Groes, at Radboud UMC, and Miranda van Hooff, at Sint Maartenskliniek and Radboud UMC, for their input on the construction of the taxonomy. 
Many thanks to Josh Ring and the Adaptive Learning Lab, in particular to Susan Verbunt-Janssen, Susanne de Mooij, and Zowi Vermeire for their valuable insights, and helping with the sample questions for the educational domain. Thanks to Cătălina Rățală and the Decision Science Group for their questions and feedback on a presentation of our paper. 
Comments from three ACM FAccT reviewers on a previous version, as well as comments from three AAAI/ACM AIES reviewers helped us to improve the paper. This research is funded by the Donders Centre for Cognition.
\end{acks}

\bibliographystyle{ACM-Reference-Format}
\bibliography{bib}

\end{document}